
%
%
%
\def\UCR{
{{\it University of California at Riverside\break
                  Department of Physics\break
                  Riverside, California 92521--0413; U.S.A. \break
                  WUDKA{\rm @}UCRPHYS}}}
\def\and{{\it\&}}
\def\half{{1\over2}}

\def\gesim{\,{\raise-3pt\hbox{$\sim$}}\!\!\!\!\!{\raise2pt\hbox{$>$}}\,}
\def\lesim{\,{\raise-3pt\hbox{$\sim$}}\!\!\!\!\!{\raise2pt\hbox{$<$}}\,}
\def\boldoverdot{\,{\raise6pt\hbox{\bf.}\!\!\!\!\>}}

\def\ibid{{\it ibid.}\ }
\def\etal{{\it et. al.}\ }
\def\acal{{\cal A}}

\def\lcal{{\cal L}}

\def\ocal{{\cal O}}

\def\vev{vacuum expectation value}

\def\diag{\hbox{\diag}}

\def\gev{\hbox{GeV}}
\def\tev{\hbox{TeV}}

%
%
%
\def\inbox#1{\vbox{\hrule\hbox{\vrule\kern5pt
     \vbox{\kern5pt#1\kern5pt}\kern5pt\vrule}\hrule}}
\def\sqr#1#2{{\vcenter{\hrule height.#2pt
      \hbox{\vrule width.#2pt height#1pt \kern#1pt
         \vrule width.#2pt}
      \hrule height.#2pt}}}

\def\today{\ifcase\month\or
  January\or February\or March\or April\or May\or June\or
  July\or August\or September\or October\or November\or December\fi
  \space\number\day, \number\year}
\def\pmb#1{\setbox0=\hbox{#1}%
  \kern-.025em\copy0\kern-\wd0
  \kern.05em\copy0\kern-\wd0
  \kern-.025em\raise.0433em\box0 }
\def\lowti#1{_{{\rm #1 }}}
\def\inv#1{{1\over#1}}
\def\su#1{{SU(#1)}}
\def\ui{U(1)}
%

%

%
\def\sumprime_#1{\setbox0=\hbox{$\scriptstyle{#1}$}
  \setbox2=\hbox{$\displaystyle{\sum}$}
  \setbox4=\hbox{${}'\mathsurround=0pt$}
  \dimen0=.5\wd0 \advance\dimen0 by-.5\wd2
  \ifdim\dimen0>0pt
  \ifdim\dimen0>\wd4 \kern\wd4 \else\kern\dimen0\fi\fi
\mathop{{\sum}'}_{\kern-\wd4 #1}}
%
%
%
%
\font\sanser=cmssq8
\font\sanseri=cmssq8 scaled\magstep1

\font\bigboldiii=cmbx10 scaled\magstep3

\def\theabstract{I review the use of effective lagrangians in describing
the physics beyond the standard model, several theoretical and practical
aspects are discussed. It is argued that the only situations where new
physics can be observed corresponds to cases where the  standard model
contributions are extremely suppressed and where high quality data is
avaliable}

\input phyzzx
\pubnum={108}
\unnumberedchapters
{\titlepage \title{ {\bigboldiii EFFECTIVE LAGRANGIANS AND TRIPLE BOSON
COUPLINGS}}
\singlespace
\bigskip
\author{{ Jos\'e Wudka}}
\address{ \UCR }
\abstract \sanseri \theabstract
\endpage}

\REF\bw{W. Buchm\"uller and D. Wyler, Nucl. Phys. B{\bf268}, 621 (1986).}
\REF\all{
M.B. Einhorn and J. Wudka, lectures presented at the {\sl Workshop on
Electroweak Symmetry Breaking}, Hiroshima, Nov. 12-15 1991, and at the {\sl
Yale Workshop on Future Colliders}, Oct. 2--3, 1992 (both presented by MBE.)
A. DeR\'ujula, \etal, Nucl. Phys. B357, 311 (1991).
J.-M. Fr\`ere \etal, preprint CERN --TH.6573/92, ULB -- TH -- 04/92.
G. Gounaris and F.M. Renard, preprint THES-TP 92/11
G. Gounaris, these proceedings.
K. Hikasa \etal, preprint MAD/PH/737.}
\REF\derujula{A. DeR\'ujula, \etal, Nucl. Phys. B357, 311 (1991).}
\REF\ndc{
T. Appelquist and C. Bernard, Phys. Rev. D{\bf22}, 200 (1980).
A.C. Longhitano, Nucl. Phys. B{\bf188}, 118 (1981).
M. Chanowitz and M.K. Gaillard, Nucl. Phys. B{\bf261}, 379 (1985).
J. Gasser and H, Leutwyler, Nucl. Phys. B{\bf250}, 465 (1985).
M. Chanowitz \etal, Phys. Rev. D{\bf36}, 1490, (1987).
J. Gasser \etal,Nucl. Phys. B{\bf307}, 779 (1988).
B. Grinstein and M. Wise, preprint HUTP-91/A015.
T. Appelquist and J. Terning, preprint YCTP-P41-92.
J. bagger \etal, preprint FERMILAB-PUB-92/75T.
T. Appelquist and G.-H. Wu, preprint YCTP-P7-93.}
\REF\geo{H. Georgi, Nucl. Phys. B{\bf361}, 339 (1991), \ibid\ {\bf363}, 301,
1991}
Effective lagrangians have been used often in the past with great sucess as,
for example, the four-fermi approach to low energy weak interactions and the
chiral lagrangian approach to the strong interactions demonstrate. In this talk
I will apply this formalism to describe the low energy effects of physics
beyond the standard model [\bw,\all,\derujula,\ndc,\geo]. This approach is
self-consistent and model independent but cannot (by its very nature) determine
unambiguously the kind of physics present at high energies. A well known fact
(though often forgotten) is that the effective lagrangian is associated with a
cutoff $ \Lambda $ which is a measure of the scale of new physics; it then
follows that this approach has a limited energy range and cannot be applied for
energies above $ \Lambda $.

\REF\dec{T. Appelquist and J. Carazzone, Phys. Rev. D{\bf11}, 2856 (1975).}
The effective lagrangian is defined as the most general (local) object which
obeys certain symmetries and contains a given set of fields (the light
excitations). For the case at hand I will choose the standard model fields
(including the Higgs) as the light excitations; in this case the decoupling
theorem [\dec] insures that all observable low effects produced by the new
interactions can be described as a series in $ 1 / \Lambda $ [\bw]. When the
Higgs is not included in the light sector a chiral lagrangian classification of
the induced operators is required and has been studied extensively elsewhere
[\ndc]. In this talk I will consider only first possibility due to space
limitations.

\REF\tbv{See for example,
K. Hagiwara \etal, Nucl. Phys. B{\bf262}, 204 (1985).
E. Yehudai, Phys. Rev. D{\bf41}, 33 (1990)
K. Hikasa \etal, Phys. Rev. D{\bf41}, 2113 (1990).
K.A. Peterson, preprint Alberta Thy-4-91.
D. London and C.P. Burgess, preprint UdeM-LPN-TH-104.
A.F. Falk \etal, Nucl. Phys. B{\bf365}, 523 (1991).
C. Grosse-Knetter and D. Schildknecht, preprint BI-TP 92/30.
J.-L. Kneur \etal, preprint CERN.TH.5979/91
G. Gounaris, ref. \all.} An example of an effective lagrangian extensively
studied in the literature [\tbv] is the triple gauge bosons vertex for two
$W$'s and one neutral ($ Z , \gamma $) vector boson, $$ \eqalign{ \lcal_{ WWV }
& = g_{ WWV}\Bigl[ i g_1^V \left( W^\dagger_{ \mu \nu } W^ \mu V^\nu - \hbox{
h. c .} \right) + i \kappa_V W_\mu^\dagger W_\nu V^{ \mu \nu } \cr & + i {
\lambda_V \over m_W^2 }W^\dagger_{ \lambda \mu } W^{ \mu \nu } V_\nu{}^\lambda
- g_4^V W_\mu^\dagger W_\nu \left( \partial ^\mu V^\nu + \partial^\nu V^\mu
\right) \cr & + \hbox{ CP \ violating \ terms } \Bigr] \cr } \eqn\lwwv $$ where
the imposed symmetries are $ U ( 1 ) _{ E M } $ (for the case $V = \gamma$) and
Lorentz invariance. This expression predicts the following static moments for
the $W$: $ \mu_W = e ( 1 + \kappa_\gamma + \lambda _\gamma ) / ( 2 m_W )$ for
the magnetic dipole moment and $ Q_W = - e ( \kappa_\gamma - \lambda_\gamma ) /
m_W^2 $ for the electric quadrupole moment; for the standard model $ g_1^\gamma
= 1, \ \lambda_{ Z , \gamma } = 0 , \ \kappa_{ Z , \gamma } = 1 $.

\REF\bl{C.P. Burgess and D. London, preprint McGill 92/04.}
\REF\stu{See, for example, S.J. gates Jr. \etal {\sl Superspace}, Benjamin
Cummings, Reading, Mass (1983) sect. 3.10.}
Going back to the general formalism I need now to choose the symmetries to be
obeyed by the effective interactions. In this respect there has been some
controversy as to whether gauge invariance should be imposed; the point [\bl]
is that, by an extension of the Stuckelberg trick [\stu], any lagrangian can be
thought as being the the unitary gauge version of a gauge invariant lagrangian.
The idea is simple: given a set of vector fields $A_\mu $, introduce an
auxiliary unitary field $U$ and construct the object $ \acal_\mu = U^\dagger (
\partial_\mu + i A_\mu ) U $; if we now assume that the $ A_\mu $ are in fact
gauge fields, then a gauge transformation for $U$ can be chosen so that $ \acal
$ is gauge covariant. Then if the original lagrangian is $ \lcal ( A ) $ the
object $ \lcal ( \acal ) $ is gauge invariant and coincides with $ \lcal ( A )
$ in the unitary limit $ U \rightarrow 1 $.

\REF\lep{See the talks by M. Vysotsky, F. Marion, J. Bourdeau, C. Declercq, E.
Lieb,, V. Innocente, E. Gross, D. Charlton and R. Barbieri in these
proceedings}
Based on this it would seem that gauge invariance is indeed a red herring
(since anything can be thought of as being gauge invariant) but I don't belive
that this is the case. The important point often missed in this discussion is
that {\it the gauge group is not fixed}. For example, $ \lcal_{ W W V } $ in
\lwwv\ can be written as an $ \su2 \times \ui $ {\it or} as a $ \ui ^3 $
invariant lagrangian, and the imposition of a given gauge group does have
non-trivial content. Since the standard model gauge symmetry is in agreement
with all experimental observations [\lep], I will also require the effective
interactions to obey this symmetry.

\REF\veltman{M. Veltman, Acta. Phys. Pol. B{\bf12}, 437 (1981)}
{}From a theoretical standpoint gauge invariance is also very important for the
naturalness of the theory: if absent, strong arguments indicate that radiative
corrections will drive the mass of the vector bosons to the cutoff [\veltman].

\REF\more{
C.J.C. Burgess and H.J. Schnitzer, Nucl. Phys. B{\bf288}, 464 (1983).
C.N. Leung \etal, Z. Phys. C{\bf31}, 433 (1986.)
W. Buchm\"uller \etal, Phys. Lett. B{\bf197}, 379 (1987).}
Based on the above discussion the effective lagrangian can be written in the
form $$ \lcal \lowti{ eff } = \inv{ \Lambda^2 } \sum_ \ocal \alpha_ \ocal \ocal
+ O \left( \inv{ \Lambda^3 } \right ) \eqn\eq $$ where catalogues of the
operators $ \ocal $ can be found in Refs. \bw, \more; some examples are $$
\ocal_W = \epsilon_{a b c } W^a_{ \mu \nu } W^b{}^{ \nu \lambda } W^c{}_\lambda
{} ^\mu ; \qquad \ocal_{ WB } = \left( \phi^\dagger \tau^a \phi \right) W^a_{
\mu \nu} B^{ \mu \nu } \eqn\eq $$ where $ W^a_{ \mu \nu } $ and $ B_{ \mu \nu }
$ denote respectively the $ \su2 $ and $ \ui $ field curvatures and $ \phi $
the scalar doublet. These operators contribute to \lwwv, for example: $
\kappa_\gamma - 1 = ( 4 m_W^2 / g g' \Lambda^2 ) \alpha_{ W B } $ and $
\lambda_ \gamma = ( 6 m_W^2 / g \Lambda^2 ) \alpha_W $.

\REF\ew{M.B. Einhorn and J. Wudka, in preparation }
The requirement that the approach be self consistent fixes the order of
magnitude of the coefficients, for example, if the underlying physics is weakly
coupled it can be shown that, since $ \ocal_W $ and $ \ocal_{ WB } $ can be
generated only via loops [\ew], $$ \alpha_W \sim { g^3 \over 16 \pi^2 } ;
\qquad \alpha_{ W B } \sim { g g' \over 16 \pi^2 } \eqn\estim $$ where I used
the fact that each vector bosons is always accompanied with its corresponding
gauge coupling. Therefore the natural size for the anomalous moments is $$ |
\kappa_\gamma - 1 | , | \lambda_\gamma | \sim 10^{ - 3 } \qquad ( \Lambda \sim
250 \gev ) \eqn\eq $$

\REF\ssc{ C. Grosse-Knetter and D. Schildknecht, ref. \tbv}
\REF\barklow{T. Barklow, lecture presented at the 1993 Aspen Winter Conference}
This is bad news for precision measurements at colliders. To see this note that
effective lagrangians are useful only if the underlying physics is not
apparent, thus if, for example, we hope to use this approach in a $1\tev$
electron collider, we should study the predictions with $ \Lambda > 1 \tev $
only. Applying this to LEP2, I find that the expected magnitude of $ | \kappa -
1 | $ and $ \lambda $ is $ 10^{ - 3 } $, while the expected sensitivity is only
$ \sim 0.1 $. The sensitivity is expected to improve by an order of magnitude
at the NLC ($ \sqrt{ s } = 0.5 {\tev} $), but then the natural scale of the
above parameters becomes $ \sim 4 \times 10^{ - 4 } $. It is, however, possible
for the underlying theory to have relatively light resonances, this in some
instances can produce an improvement of $ \sim 10 $ in the estimates for the $
\alpha_\ocal $; if this is actually realized several collider proposals to
measure $ \kappa $ and $ \lambda $ will be marginally sensitive to new physics,
some examples are the SSC (sensitivity $ \sim 10^{ - 2 } $) [\ssc] and a $ 1.5
{\tev} $ $ e^+ e^- $ collider (sensitivity $ \sim 10^{ - 3 } $) [\barklow].

\REF\peccei{ See for example K. Hagiwara \etal, ref. \tbv}
\REF\du{ C. Ahn \etal, Nucl. Phys. B{\bf309}, 221 (1988). }
A conclusion which can be immediately drawn is that there are {\it never} large
effects from $ \lcal \lowti{ eff} $. This results from the suppression by
powers of $ 1/ \Lambda $ multiplying the $ \ocal $ and from the limited energy
range of applicability. The existing claims to the contrary correspond to
situations where the energies are larger than the cutoff [\peccei]
\foot{{\sanser This is also true for the ``delayed unitarity'' scenario
[\du].}}, or from unnaturally large coefficients which violate \estim\ by
several orders of magnitude. Therefore the best places to find effects from the
operators $ \ocal $ is either in high precision experiments (such as the muon's
anomalous magnetic moment) or in cases where the standard model contributions
are accidently suppressed (such as the $ \rho $ parameter).

\REF\gmt{ C. Arzt, \etal UCRHEP-T98, see also R. Escribano and E. Mass\'o,
Phys. Lett. B{\bf301}, 419 (1993). Previous work is reviewed in T. Kinoshita
and W.J. Marciano, preprint BNL-44467}
\REF\brook{ V.W. Hughes, AIP Conf. Proceedings No. 187, 326 (1989). M. May, AIP
Conf. Proceedings No. 176, 31168 (1988).}
I shall present two examples corresponding to each of these possibilities.
First I present the results for the anomalous moment of the muon [\gmt]. I will
consider two contributions, first from the operators $ \ocal_W $ and $ \ocal_{
WB } $ at the one loop level, and the tree level ones generated by the
``direct'' operators $$ \ocal_ { \mu B } = ( \bar \nu_\mu , \bar \mu)_L
\sigma^{ \mu \nu } B_{ \mu \nu } \mu_R \phi \qquad \ocal_ { \mu W } = ( \bar
\nu_\mu , \bar \mu)_L \sigma^{ \mu \nu } \sigma^a W^a_{ \mu \nu } \mu_R \phi
\eqn\eq $$ The loop contributions give \def\ltv{\Lambda^2\lowti{TeV}} $$
\eqalign{ \delta a_\mu =& 10^{ - 10 } { \alpha_W \over \ltv } + 2 \times 10^{ -
9 } \left(1 + { \ln \ltv \over 7 } \right) {\alpha \over \ltv } \cr \simeq & 3
\times 10^{ - 12 } \left(1 + { \ln \ltv \over 7 } \right) \inv{ \ltv } \cr }
\eqn\eq $$ where $ \ltv $ is the cutoff in {\tev}\ units and I used \estim. It
is clear from this expression that the Brookhaven experiment AGS 821 [\brook]
will be insensitive to $ \ocal_W ,\ \ocal_{ WB } $, the corresponding
contributions are just too small. On the other hand, the effects of $ \ocal_{
\mu W, B } $ are much larger and in fact a sensitivity to $ \Lambda $ of $ \sim
700 \gev $ can be inferred [\gmt] for the above experiment.

\REF\hhg{J.F. Gunion \etal, {\sl The Higgs Hunters Guide}, Benjamin Cummings}
\REF\std{
G. Keller and D. Wyler, Nucl. Phys. B{\bf274}, 410 (1986).
R. DEcker \etal, Phys. Lett. B{\bf255}, 605 (1991).}
As the second example I consider a modified standard model in which the low
energy fields are the standard model ones with the addition of an extra scalar
doublet \foot{{\sanser I will impose the usual discrete symmetry [\hhg].}}.
Within this model there is a scalar excitation which is CP odd and which I'll
denote by $a_o$; its decay into two photons is strongly suppressed in the limit
where the ration of the $\vev s$, denoted by $ \tan \beta $, is large [\std].
Therefore this is a promising decay to study when considering the effects of
the effective operators $ \ocal $ (which describe now the physics beyond this
two doublet model).

\REF\maperez{ M.A. Perez J. Toscano and J. Wudka, in preparation}
There is no dimension six operator contributing to $ a_o \rightarrow \gamma
\gamma $ at tree level. so that the $ \ocal $-induced loop contributions must
be finite. This is indeed verified by explicit computation [\maperez]. For
simplicity I will consider here only $$ \ocal = ( \phi_1^\dagger \phi_2 ) (
\bar q t_R \phi_1 ) ,\eqn\eq , $$ where $q$ denotes the top-bottom left-handed
fermion doublet and $t_R$ the right-handed top quark field. The full
calculation will appear in Ref. \maperez.

Evaluating the relevant graphs and taking $ m \lowti{ top } = 170 \gev $ and $
\alpha_\ocal v^2 = \Lambda^2 $ I find that the contribution from this operator
dominates the usual one provided $ \tan \beta > 6 $, the general order of
magnitude is $ \Gamma ( a_o \rightarrow \gamma\gamma ) \sim 10^{ - 9 } \gev $
when $ m_{ a_o } \sim m \lowti{ top } $ and it peaks at $ m_{ a_o } \simeq 2 m
\lowti{ top } $. It is worth pointing out that the effective operators will
also generate angular distributions for the two photon final state which
differs from the standard one. The branching ratio is very small, however, and
this will probably be unobservable. This reaction is a good place to look for
physics beyond, for example, the minimal supersymmetric extension of the
standard model [\hhg] if the large $ \tan \beta $ scenario is realized.

I now discuss some technical aspects of the effective lagrangian formalism.

\REF\ar{C. Arzt, preprint UM-TH-92-28}
$i)$ \undertext{Equations of motion.} It has been shown that the number of
operators $ \ocal $ can be reduced when the classical equations of motion are
imposed [\bw,\geo,\ar]. However, even if some operators are related via the
classical equations of motion, they can have very different origins. Consider
for example $$ \left\{ ( D_\mu \phi)^\dagger ( D_\mu \phi ) + \phi^\dagger [
D_\mu , D_\nu ] \phi \right\} B^{ \nu \mu } \eqn\eq $$ which can be generated
only via loops [\ew]. On the other hand the use of the equations of motion show
that it is equivalent to $ ( \phi^\dagger D_\nu \phi) j^\nu $ where $j^\nu $ is
the source current for $B$; this last operator can be generated at {\it tree}
level. Therefore even if the S-matrix elements cannot distinguish between the
first and second operators, there is a very large quantitative difference
whether the underlying physics generates the second one or not.

$ii)$ \undertext{Blind directions} The above comments should be kept in mind
when studying effects from operators to which we are not currently sensitive
(blind directions [\derujula]). In the final analysis, the statement that blind
operators have coefficients similar to the ones we are sensitive to is an
additional assumption. To illustrate this point consider a model with a light
scalar field $ \phi $ interacting with two heavy fermions $ \psi_a \ (a = 1 ,2
)$. The lagrangian is $$ \lcal = \underbrace{ \half ( \partial \phi )^2 - \half
m^2 \phi^2 - \inv6 \sigma \phi^3 - \inv{24} \lambda \phi^4 }_{\hbox{{\sanser
light \ sector }}} + \underbrace{ \sum_{ a = 1 }^2 \bar \psi_a \left( i \not
\partial - M + ( - )^a g \phi \right ) \psi }_ {\hbox{{\sanser heavy \ sector
}}} \eqn\eq $$ When the fermions are integrated out they produce an effective
action even in $ \phi $.

If odd powers of $ \phi $ are blind in this toy world then the claim that, say,
the coefficient of $ \phi^5 $ will be of the same order as that of $ \phi^6 $
(times $M$) is wrong. On the other hand if even powers of $ \phi $ are blind,
then very precise measurements on the coefficient of $ \phi^5 $ will produce
very misleading conclusions about the scale $M$.

\REF\anom{J. Minn, \etal, Phys. Rev. D{\bf35}, 1872 (1987).} [\anom]
$iii)$ \undertext{Anomalies} Though there are new fermionic couplings no new
anomalies are generated

\REF\bg{
B.S. DeWitt, in ``Quantum Gravity 2'', edited by C.J. Isham, R. Penrose and
D.W. Sciama.
G. t'Hooft, in ``Karpacz 1975, Proceedings'', Acta Universitatis
Wratislaviensis, No. 368, Vol. 1, (Wroclaw 1976), pp. 345--369.
L.F. Abbott, Acta Phys.\ Pol.\ {\bf B13}, 33 (1982).
The modification for spontaneously broken gauge theories is given in M.B.
Einhorn and J. Wudka, Phys.\ Rev.\ D{\bf39}, 2758 (1989).}
$iv)$ \undertext{Gauge invariance} It might seem puzzling to assume, from a
calculational point of view, that the effective operators are gauge invariant.
After all these are generated in a large mass expansion from an underlying
(presumably gauge) theory. In the process one must fix the (underlying) gauge,
add Fadeev-Popov ghost, etc. etc. and in this process explicit gauge invariance
is generally lost. To solve this puzzle it is just necessary to recall that one
can use a background gauge fixing method [\bg] which guarantees a gauge
invariant effective action.

\REF\loop{ For the decoupling case see for example C. Arzt \etal, ref. \gmt,
C.P. Burgess and D. London, preprints McGill 92/05, 92/38.
M.-A. Perez \etal, Phys. Lett B{\bf289}, 381 (1992)
P. Hernandez and F.J. Vegas, preprint CERN.TH 66670.
For the non-decoupling case see refs. \ndc.}
$v)$ \undertext{Renormalizability}. Contrary to the standard lore, effective
lagrangians {\it are} renormalizable: all infinities can be absorbed in
redefining the coefficients of some effective operator already present (by
definition) in the theory, moreover this can be done order-by-order in $ 1 /
\Lambda $ . The only remnant of the (logarithmic) divergences is the
renormalization group running of the coefficients $ \alpha_\ocal $. All power
divergences are unobservable. Many loop calculations have already appeared in
the literature [\loop].

I would like to conclude by stating that the effective lagrangian formalism is
a self-consistent model independent way of talking about what we don't know.
The approach is fully renormalizable and the divergences are unobservable,
loops can be calculated as usual.

\REF\frere{J.-M. Fr\`ere \etal, ref. \all}
Most effects are predicted to be too small for observation and the expected
sensitivity to be derived from near-future and existing experiments is modest
(in the few hundred \gev\ range). It is possible, however, for the coefficients
$ \alpha_\ocal $ to be anomalously small [\frere], in this case it may very
well be that no precision measurement will suggest the presence of new
resonances which, in fact, are just around the corner. There is also the
possibility of some modest ($ \lesim 10 $) enhancements of the $ \alpha_\ocal $
due, for example, to low lying resonances, but even in this case the
sensitivity of future colliders will be at best marginal to the effects of the
operators $ \ocal $

\refout

\bye